\documentclass[conference]{IEEEtran}
\IEEEoverridecommandlockouts

\usepackage[utf8]{inputenc}
\IEEEoverridecommandlockouts
\usepackage{cite}
\usepackage{amsmath,amssymb,amsfonts}
\usepackage{algorithmic}
\usepackage{graphicx}
\usepackage{textcomp}
\usepackage{xcolor}
\usepackage{multirow}
\usepackage{tabularx}
\usepackage{todonotes}
\usepackage{array}
\newcolumntype{?}{!{\vrule width 1pt}}
\usepackage{float}
\newcolumntype{C}[1]{>{\centering\arraybackslash}p{#1}}
\usepackage{subcaption}
\def\BibTeX{{\rm B\kern-.05em{\sc i\kern-.025em b}\kern-.08em
    T\kern-.1667em\lower.7ex\hbox{E}\kern-.125emX}}

\begin{document}

\title{Towards a Holistic Definition of Requirements Debt}

\author{
\IEEEauthorblockN{Valentina Lenarduzzi}
\IEEEauthorblockA{
\textit{Tampere University}\\
Tampere, Finland \\
valentina.lenarduzzi@tuni.fi}
\and
\IEEEauthorblockN{Davide Fucci}
\IEEEauthorblockA{
\textit{University of Hamburg}\\
Hamburg, Germany \\
fucci@informatik.uni-hamburg.de}
}




\maketitle
\begin{abstract}
When not appropriately managed, technical debt is considered to have negative effects to the long term success of software projects.
However, how the debt metaphor applies to requirements engineering in general, and to  requirements engineering activities in particular, is not well understood. 
Grounded in the existing literature, we present a holistic definition of requirements debt which includes debt incurred during the identification, formalization, and implementation of requirements. 
We outline future assessment to validate and further refine our proposed definition.
This conceptualization is a first step towards a requirements debt monitoring framework to support stakeholders decisions, such as when to incur and eventually pay back requirements debt, and at what costs. \\
\end{abstract}


\section{Introduction}
\label{Introduction}

Cunningham defines Technical Debt (TD) as \textit{''The debt incurred through the speeding up of software project development which results in a number of deficiencies ending up in high maintenance overheads''}~\cite{Cunningham1992}. 
TD implies sub-optimal design or implementation solutions giving a short-term benefit while making changes more costly or even impossible in the medium and long term.
Unpredictable business and environmental forces, internal or external to a company, can result in TD which needs to be managed~\cite{Martini2015, Besker2018}. 

The TD metaphor was initially concerned with software implementation (i.e., at code level), but it has been gradually extended to software architecture, design, documentation, testing, and requirements~\cite{Brown2010}.
Li et al.~\cite{Li2015} conducted a systematic mapping study on understanding and managing TD drawing an overview on the current state of research.
They proposed a classification of 10 types of TD: Requirements, Architectural, Design, Code, Test, Build, Documentation, Infrastructure, and Versioning. 

Requirements Debt  (ReD) ''\textit{refers to the distance between the optimal requirements specification and the actual system implementation, under domain assumptions and constraints}''~\cite{Ernst2012}.
Despite the importance of requirements engineering activities during software development process~\cite{Schmid2013, Alves2014} and the definition of the minimum viable product (MVP)~\cite{LenarduzziMVP}, there is still no consensus in research whether ReD should be considered as a type of technical debt or not~\cite{Alves2014}.
Different processes could led to different requirement decomposition and accumulate different debt~\cite{TaibiXP2017}.
We believe the reason is the lack of formalization of ReD in the literature~\cite{Li2015},~\cite{LenarduzziSLRTD2019}.
 
In this paper, we challenge the current definition of ReD~\cite{Ernst2012}---which focuses on downstream activities in the software development lifecycle, such as software development and evolution---and extend it to include upstream activities involving the elicitation of requirements (particularly in user-centered requirements engineering~\cite{MNJ15}) and their translation into specifications.



Our definition of ReD is the first step towards creating a framework that stakeholders can use to make decisions regarding when to incur debt, at what costs, when to pay it back, and how to monitor it. 
Our vision of ReD will be empirically evaluated in a series of studies with industry partners and individual stakeholders. 


In this paper, we aim at providing a holistic definition of ReD which takes into account the relevant   requirements engineering activities currently investigated in the literature. 
Moreover, we outline the future assessments to conceptualize and define the decision framework. 


\section{Definitions}
\label{Background}
In this section, we report the basic definitions of concepts associated to Technical Debt (TD) used to define Requirement Debt (ReD).

\vspace{2mm}
\noindent\textit{Principal}. In finance, it refers to the original amount of money borrowed.
From a software development perspective, the term is used to describe the cost of remediating planned software system violations. 
Ampatzoglou et al.~\cite{AMPATZOGLOU2015} defines principal within a TD context as: ``The effort that is required to address the difference between the current and the optimal level of design-time quality, in an immature software artifact or the complete software system.''

\vspace{2mm}
\noindent\textit{Interest}. It is the negative effects of the extra effort that has to be paid due to the accumulated amount of debt in the system, such as executing manual processes that could potentially be automated, excessive effort spent on modifying unnecessarily complex code, performance problems due to lower resource usage by inefficient code, and similar costs~\cite{TOM2013}. 
Ampatzoglou et al.~\cite{AMPATZOGLOU2015} defines interest as: ``The additional effort that is needed to be spent on maintaining the software, because of its decayed design-time quality.''

\vspace{2mm}
\noindent\textit{Quantify and paying back TD}. The principal should be paid if it is less than the total interest~\cite{Seaman2011}. 
Refactoring decision depends on the ratio between Principal and Interest. 
If the value is greater than one, it is not convenient to pay the principal now with respect to the interest that will be paid in the future~\cite{Martini2016}. 









\section{ReD---Requirements Debt}
\label{ReD}
In this section, we provide a definition of Requirements Debt (ReD), propose its conceptualization, and strategies for detecting, quantifying, and paying it back.

\begin{figure} [H]
    \centering
    \includegraphics[trim={8cm 5cm 8cm 4cm}, clip]{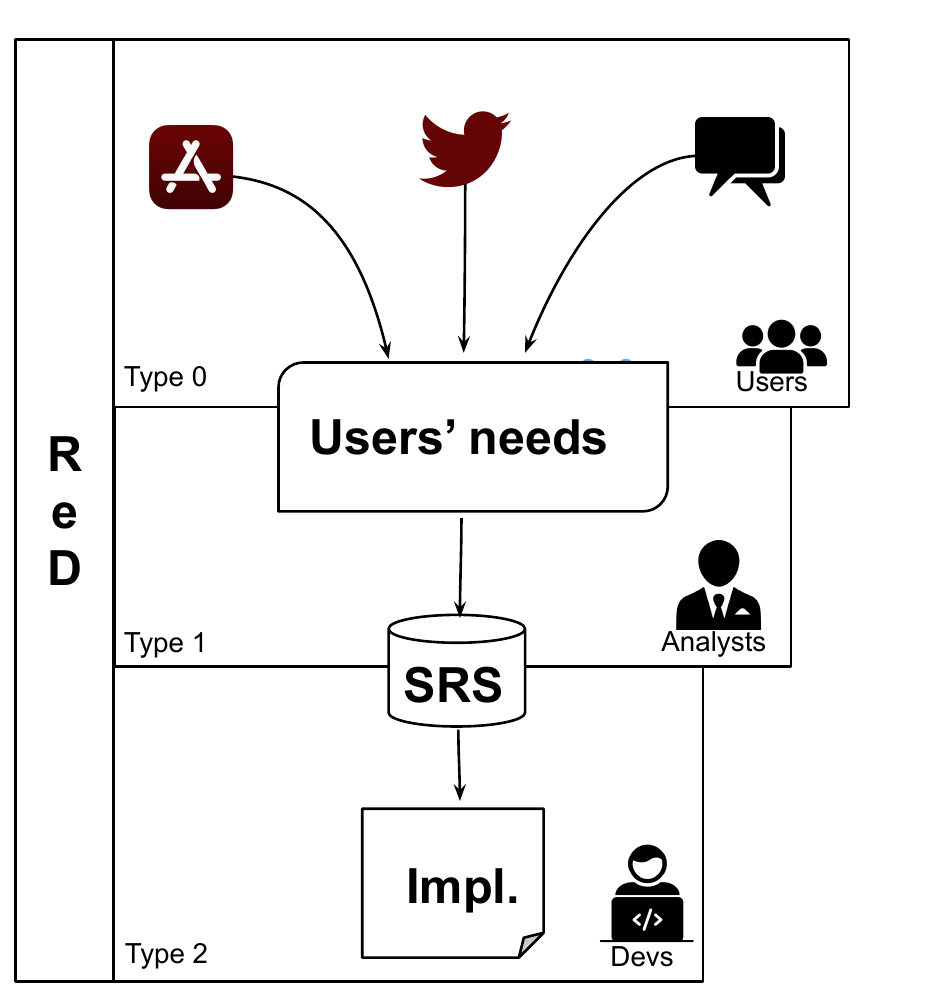}
    \caption{Types of Requirements Debt (ReD) incurred, their relationship, and main stakeholder involved.}
    \label{fig:red}
\end{figure}

We define three types of ReD (see Figure~\ref{fig:red}).

\subsection{ReD Type 0: Incomplete Users’ needs} 
Represents the debt incurred when neglecting users' needs expressed using feedback channels. 
Such channels are, for example, app stores, social media, and interviews with customers. 
In particular, this type of debt is incurred when i) within a channel, not all users' needs are captured (e.g., due to the complexity of processing large amount of feedback) and ii) one or more relevant channels are not considered.
In both cases, the incurred ReD can be implicit (e.g., causing an unplanned cost) or explicit (e.g., necessary due to a deadline). 


 \vspace{2mm}
 \noindent\textbf{How to detect.} Currently, i) could be addressed by leveraging techniques for automatically classifying users' feedback and stakeholders, summarizing it, and recommend new features based on it. One example is the work of Maalej et al. which applies such techniques for app stores analytics~\cite{MNJ15}.
 Regarding ii) Nayebi et al.~\cite{NHR18} presents the limitations of app stores in acquiring users' needs. The authors show that Twitter can provide additional information which app developers can exploit to deliver better products. Moreover, feedback reported in Twitter is more objective than the one reported in app reviews. This work shows the importance of investigating several sources of users' needs and presents an initial approach to detect whether all relevant channels are considered.
 
 \vspace{2mm}
 \noindent\textbf{How to quantify.} ReD Type 0 can be quantified as the cost to formalize and implement the neglecting needs (\textit{Principal}).  We should considered two extra extra effort (\textit{Interest}) related to:
 \begin{itemize}
     \item  the current development stage, as implementing a neglected need is more expensive in an advanced stage,
     \item  which components need to be modified to fix ReD. For example, implementing a neglected users' need impacting the graphic user interface will cost less than one impacting the architecture. 
 \end{itemize}

\noindent\textbf{How to pay back.} Once the neglected users' need is identified, it is formalized and included in the software requirements specification document. 

\subsection{ReD Type 1: Requirement smells} 
Represents the debt incurred when a requirements engineer, business analyst, or developer (i.e., \textit{analyst}) formalizes users' needed into SRS. 
Femmer et al.~\cite{FFW17} defined a set of Requirements Smells (Table~\ref{tab:ReDSmells})---i.e., linguistic constructs which can indicate a violation of the ISO29148 standard for requirements quality.
If such ambiguity is not removed the requirement can be wrongly implemented, hard to reuse, evaluate and extend. 

\begin{table*}[t]
\centering
\caption{Requirement Smells~\cite{FFW17}}
\label{tab:ReDSmells}
\begin{tabular}{p{4.2cm}|p{10cm} p{2.5cm}} \hline 
\textbf{Requirement Smells} & \textbf{Description} & \textbf{Detection Strategy}\\ \hline 
Subjective Language & ''Subjective Language refers to words of which the semantics is not objectively defined, such as user friendly, easy to use, cost effective'' & Dictionary  \\\hline 
Ambiguous Adverbs and Adjectives  & ''Ambiguous Adverbs and Adjectives refer to certain adverbs and adjectives that are unspecific by nature, such as almost always, significant and minimal'' & Dictionary\\\hline 
Loopholes & ''Loopholes refer to phrases that express that the following requirement must be fulfilled only to a certain, imprecisely defined extent'' & Dictionary \\\hline 
Open-ended, non-verifiable terms  & ''Open-ended, non-verifiable terms are hard to verify as they offer a choice of possibilities, e.g. for the developers'' & Dictionary\\\hline 
Superlatives & ''Superlatives refer to requirements that express a relation of the system to all other systems'' & Morphological Analysis, POS tagging \\\hline 
Comparatives & ''Comparatives are used in requirements that express a relation of the system to specific other systems or previous situations'' & Morphological Analysis, POS tagging \\\hline 
Negative Statements & ''Negative Statements are statements of system capability not to be provided. Some argue that negative statements can lead to under specification, such as lack of explaining the system’s reaction on such a case'' & Dictionary, POS tagging\\\hline 
Vague Pronouns & ''Vague Pronouns are unclear relations of a pronoun'' & POS tagging\\ \hline 
\end{tabular}
\end{table*}

\vspace{2mm}
\noindent\textbf{How to detect.}
Like code smells, requirements smells do not necessarily lead to a defect, can be (semi-)automatically detected within a SRS document, and removed using standard techniques.
These techniques leverage natural language processing (e.g., POS tagging) or simple dictionary lookups to identify problematic terms and language constructs (Table~\ref{tab:ReDSmells}).


\vspace{2mm}
\noindent\textbf{How to quantify.}  ReD Type 1 can be quantified as the cost to fix the requirement smells within a SRS (\textit{Principal}). 
However, the cost related to the harmfulness of each requirement smell (\textit{Interest}) needs to be considered.
With harmfulness, we mean the different negative impact that each requirements smell can have on activities relying on SRS. 

\vspace{2mm}
\noindent\textbf{How to pay back.} As for code smells, refactoring (e.g., removing a problematic language construct leading to ambiguity while maintaining the original goal of the specification) is needs to be applied to pay back this type of ReD. 

\subsection{ReD Type 2: Mismatch implementation} 
Represents the debt incurred when developers implement a solution to a requirement problem. 
This type captures the mismatch between stakeholders' goal framed in the SRS and the actual system implementation.  
This type of debt can be also incurred when the requirements problem, framed in the SRS, changes while the implementation does not change accordingly~\cite{Ernst2012}. 
A sub-par implementation can be the result of the incurred Type 1 debt.

\vspace{2mm}
\noindent\textbf{How to detect.} Detection of this type of ReD can be based on approaches for traceability between SRS and source code. 
Knowledge-based approaches (e.g., \texttt{RE-KOMBINE}~\cite{Ernst2012}) can be used to monitor requirements for changes and understanding their impact on the current implementation of the system. 

\vspace{2mm}
\noindent\textbf{How to quantify.} The interest on ReD Type 2 is the amount of change between the current implementation and the SRS. Accordingly, it is quantified as the cost of comparing the current implementation with the set of possible changes~\cite{Ernst2015} (\textit{Principal}) plus the implementation of the selected change (\textit{Interest}).

\noindent\textbf{How to pay back.} The actions for paying back this type of ReD consists in the implementation of the best new solution matching the updated SRS. 
    


\section{Future Assessment}
\label{Future}
\subsection{ReD concept preliminary validation}
Our next step is to preliminary validate the conceptualization of debt in requirements engineering.
To that end, we designed a study to understand what practitioners consider as debt during requirements engineering process, compare it to the ReD conceptualization, and understand their motivations to incur the different types of ReD.



\begin{figure}[H]
    \centering
     \includegraphics[trim={5cm 7cm 7cm 7cm},clip, width=0.5\textwidth]{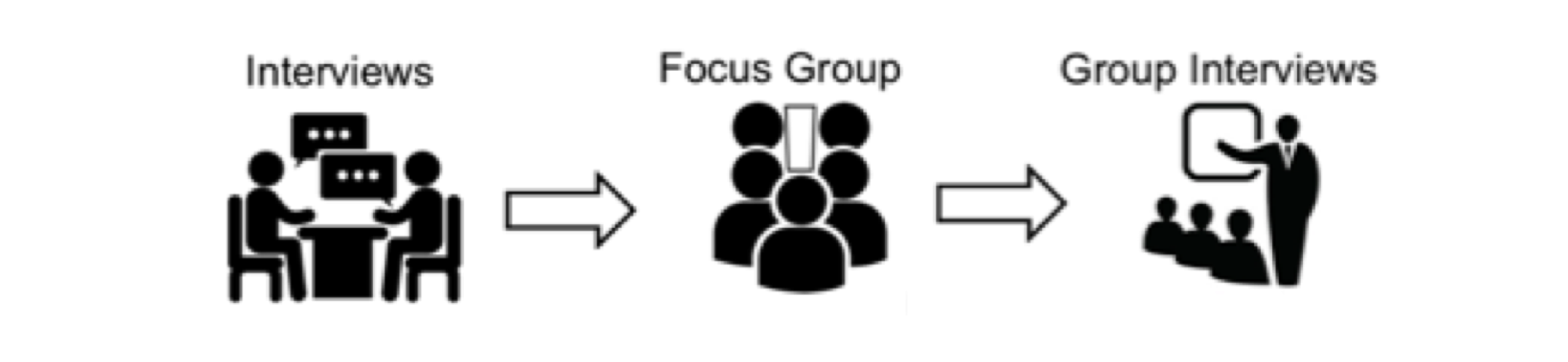}
    \caption{Study design for ReD concept validation.}
    \label{fig:surveysedign}
\end{figure}

We will carry out an exploratory study (Figure \ref{fig:surveysedign}), structured as a mixed research method, composed by a set of interviews, a focus group, and a final set of group interviews.

Based on these results, we will design and conduct detailed case studies, involving companies in order to monitor the requirement elicitation process. 

\textbf{Interviews.} The first round of interviews will be carried out by means of a questionnaire based on open-ended questions to avoid driving the interviewee to a predefined set of answers. 
The interviews will be organize in three sections.

\vspace{2mm}
\noindent \textit{1) Personal and company information.} We aim to collected the profile of the practitioners,  considering age, country, gender, predominant roles, and working experience in requirement engineering. Moreover, we will collect the organization size via the number of employees and the common application domain.

\vspace{2mm}
\noindent \textit{2) Requirement Debt (ReD)}. We will include questions regarding the three Requirements Debt types (Type 0, 1 and 2) as defined in Section~\ref{ReD}. We will ask the practitioners to evaluate and discuss these definitions from their point of view. 




\vspace{2mm}
\noindent \textit{3) Perceived Critically of Requirement Dept Types}. We aim to capture the perception of requirements issues from our respondents. We will ask practitioners to rate their concerns about requirement debt and what they consider harmful. 


\vspace{2mm}
\textbf{Focus group}. Relevant issues and problems are freely discussed and the answers provided during the interviews will cluster into topics. Open discussion can reveal the type of information that can be helpful in outlining key issues in each of the three types of ReD. 

\vspace{2mm}
\textbf{Group interviews.} The last step will be executed with the support of a closed-ended questionnaire, based on the clustered answers identified in the focus group.  The interviewer will explain each question to the participants who answered to the questions on a paper-based questionnaire.
The interviews will be organized considering only the questions related to \textit{Requirement Debt (ReD)} and \textit{Perceived Critically of Requirement issues}. 

\vspace{2mm}
\noindent \textbf{Recruitment and Data Collection}. The study will be conducted by invitation only to have a better control over the individual respondents.
The strategy to define an invitation list is two-fold, i) requirements engineers, business analysts, and software developers within companies in the \texttt{NaPiRE} network~\cite{FWK17} and among our contacts, and ii) software developers sampled from the app stores (e.g., Google Play Store, Apple Store). 
The latter are especially important to assess Type 0 ReD as they usually have direct access to users through the stores feedback and review functionalities. 

\subsection{Requirement smells harmfulness}
In the immediate future, we plan to elaborate on Type 1 ReD as there is no empirical evidence of harmfulness of requirement smells, according to the definition and the detection approach proposed by Femmer et al.~\cite{FFW17}.
We will follow the approaches widely adopted to assess the harmfulness of code smells on different software qualities~\cite{Olbrich2010}, \cite{Sjoberg2013}, \cite{Hall2014}, \cite{Palomba2018}.
We will triangulate data from requirements platforms, such as issue trackers and requirements repositories platforms~\cite{DataSet}, with studies involving requirements engineers, business analysts, and software developers~\cite{TaibiIST}. 

\section{Related Work}
\label{RW}
In this section, we report key related work on Technical Debt (TD) and Requirement Technical Debt (ReD) evaluation and management.

\subsection{Technical Debt}
Different approaches and strategies have been proposed to evaluate TD. 
Nugroho et al.~\cite{Nugroho2011} proposed an approach to quantify debts in terms of cost to fix technical issues and its interest. They monitored data from 44 software systems and empirically validated the approach in a real system.

Seaman et al.~\cite{Seaman2011} proposed a TD management framework that formalizes the relationship between cost and benefit in order to improve software quality and help decision making process during maintenance activities.


Zazworka et al.~\cite{Zazworka2014} investigated source code analysis techniques and tools to identify code debt in software systems, focusing on TD interest and TD impact on increasing defect- and change-proneness.
They applied four TD identification techniques (code smells, automatic static analysis issues, grime buildup, and modularity violations) on 13 versions of the Apache Hadoop open source software project. They collected different metrics, such as code smells and code violations. The results showed good correlation between some metrics and defect and change proneness, such as Dispersed Coupling and modularity violations. 


Different approaches or strategies have been proposed to manage TD. 
Guo et al.~\cite{Guo2011} proposed a portfolio approach in order to help the software manager in decision making. This approach provides a new perspective for TD management.

Nord et al.~\cite{Nord2012} defined a measurement-based approach to develop metrics in order to strategically managing TD. This approach could optimize the development cost over time without stopping the development process. They successfully applied the approach to an ongoing system development effort.


\subsection{Related Work on Requirement Technical Debt}
At the best of our knowledge, there are few works that investigated Requirement Technical Debt proposing some approach to evaluate and take under control this type of debt. 

Ernst et al.~\cite{Ernst2012} defined technical debt in requirements as ''\textit{the distance between the implementation and the actual state of the world}'' They conceptualized a requirements modeling tool, \texttt{RE-KOMBINE}, that 1) identifies technical debt by means of the notion of optimal solutions to a requirements and 2) allows to understand how implementations match stakeholder goals. 

Abad and Ruhe~\cite{Abad2015} defined a systematic method to manage requirements-related decisions. The methods includes several factors that affect Technical Debt; the authors extend the concept to requirements and use historical project data to provide a predictive model for requirements decisions with the goal of reducing uncertainty.

Moreover, Wattanakriengkrai et al.~\cite{Wattanakriengkrai2018} investigated self-admitted requirement debt---defined as ''\textit{source code comments deliberately created by developers in order to demonstrate that some parts of the code are missing, incomplete, or cannot satisfy the requirement of clients}''---that developers clearly identify in the code due to requirements incompleteness.
Based on this definition, they show an approach to identify requirement self-admitted technical debt on 10 open source projects analyzed using text processing techniques. 

\section{Conclusion}
\label{Conclusion}
Despite the importance of requirements elicitation and management during software development process, there is still no consensus in research whether Requirement Debt  should be considered as a type of technical debt and a lack of formalization in the literature. 

In this paper, we challenge the current definition of ReD, extending it with upstream requirements engineering activities involving the elicitation of requirements and their translation into specifications.

Our definition of ReD is the first step towards creating a framework that stakeholders can use to make decisions regarding when to incur debt, at what costs, when to pay it back, and how to monitor it. 

Our vision of ReD will be empirically evaluated in a series of studies with industry partners and individual stakeholders. 

\section*{Acknowledgments}

\bibliographystyle{IEEEtran}
\bibliography{references.bib}

\end{document}